# Speaker Recognition – Wavelet Packet Based Multiresolution Feature Extraction Approach

Authors: Saurabh Bhardwaj, Smriti Srivastava, Abhishek Bhandari, Krit Gupta, Hitesh Bahl and J.R.P. Gupta

## Abstract

This paper proposes a novel Wavelet Packet based feature extraction approach for the task of text independent speaker recognition. The features are extracted by using the combination of Mel Frequency Cepstral Coefficient (MFCC) and Wavelet Packet Transform (WPT).Hybrid Features technique uses the advantage of human ear simulation offered by MFCC combining it with multi-resolution property and noise robustness of WPT. To check the validity of the proposed approach for the text independent speaker identification and verification we have used the Gaussian Mixture Model (GMM) and Hidden Markov Model (HMM) respectively as the classifiers. The proposed paradigm is tested on voxforge speech corpus and CSTR US KED Timit database. The paradigm is also evaluated after adding standard noise signal at different level of SNRs for evaluating the noise robustness. Experimental results show that better results are achieved for the tasks of both speaker identification as well as speaker verification.

## 1 Introduction

Human beings possess several innate biological traits that assist them in distinguishing from one another. These distinguishing traits are measurable physiological and behavioral characteristics that can be utilized to verify the identity of an individual. Biometrics can be defined as the science of measuring and statistically analyzing biological data devoted for automated identification of individuals on the basis of unique biological traits. Over the years, biometrics has emerged as the science which assimilates and tries to mimic the powers of the human brain by capturing unique personal features based on retinal or iris scanning, voice patterns, dynamic signatures, fingerprints, face recognition, or hand measurements etc. and consequently performing the task of human identification. Voice as a biometric characteristic have attracted plethora of researchers as it can be easily intercepted, recorded and processed. Moreover, voice biometrics offers simple and secure mode of remote access transactions over telecommunication networks by authenticating the speaker first and then carrying out the required transactions. Speech signal in its most generic form coveys the phonetic information related to utterance that the speaker tries to convey to the listener, information about the speaker's gender, the language being spoken, speaker's emotions and the identity of the speaker. Any speaker's identity is associated with his/her behavioral and physiological characteristics of vocal tract which are unique anatomical structures. In addition to these physical differences, each speaker has his or her unique manner of speaking, including the use of a particular accent, rhythm, intonation style, pronunciation pattern, choice of vocabulary and so on.
The applications of speech processing technology are primarily classified as: *Speech Recognition and Speaker Recognition*. Speech Recognition is the ability to identify the spoken words while Speaker Recognition is the ability to identify speaker on the basis of his/her voice characteristics [1] . Speaker recognition is further dissected into two categories, speaker verification and speaker identification. Speaker verification is the process of validating the claim of identity by a speaker consequently this type of decision is binary i.e. true or false. Whereas in speaker identification there is no prior claim of an identity, the system classifies the input test speech signal belongs to which one of the 'N' reference speakers. Speaker identification stated above is labeled as "closed-set" speaker identification which is different from "open-set" identification, as in the case of open-set, a possibility exists that the test speech signal may not belong to any of the 'N' reference speakers, hence in that case N+1 decisions exist, containing an additional result of the test signal not appertaining to any of the N reference speakers [2]. Recognition systems are also classified as *text-dependent* and *text-independent*. Some systems behave in a text-dependent way, i.e. the user utters a predefined key password from a small set of vocabulary for e.g. in telephone-based services and access control. But, text dependent type of recognition process is only feasible with "cooperative speakers". Consider criminal investigation as an application (an unwilling speaker who does not wish to get recognized), here recognition can only be performed in text-independent mode. In text independent mode there is no constraint on the words which are being spoken and thus it is more flexible where the speaker might not be aware or not willing to cooperate. Text independent speaker recognition is more challenging of the two tasks because of the use of impromptu speech signals for testing [3].Classification of speaker recognition system is depicted in fig. 1.

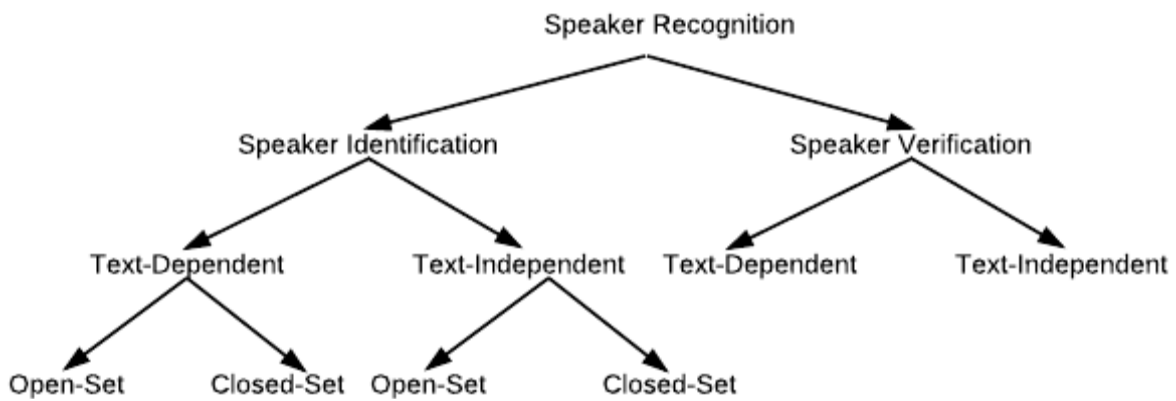

Fig.1 Classification of speaker recognition systems

With increased applications of speech as a means of communication between the man and the machine, speaker recognition has emerged as a powerful tool [4]. The phenomenon of speaker recognition has been in application since the 1970's [5]. In depth discussions of feature selection and stochastic modeling is included in the tutorial [1]. A more recent overview is provided in the tutorial [6]. Speaker recognition systems find wide variety of applications in forensics, telephone banking services, voice mail, access to information and premises to only authorized personals, in military applications to identify the suspect among set of other people using voice data [7]. Most of the state-of-the-art recognition systems uses Mel Frequency Cepstral Coefficients (MFCC) as feature extraction technique for front-end-processing as its performance is far superior compared to all other feature extraction mechanisms as described in [8]. Some of the previous works based on wavelet transform and wavelet packet transform for the task of speaker recognition are [9], [10]. [11], [12] describes some paradigms for improving the MFCC techniques.

Section 2 gives the brief descriptions of modules of speaker recognition system and feature extraction. The basic theory of the used classifier is explained in section 3. Section 4 deals with the motivation of evolving the proposed paradigm and the advantages it offer. The method is further detailed and discussed in section 5. Results are presented in section 6. Section 7 concludes the proposed research.

# 2 Modules for Speaker Recognition

All speaker recognition systems contain two main modules, *feature extraction* and *feature or pattern matching*. Feature extraction is the process that reduces the dimensionality of the input signal by deducing information related to the speaker from the voice signal. Feature matching is the process of recognizing an unknown speaker by matching its voice model with the voice models of the known speakers. Sound pressure waves are acquired with the help of a microphone. The recorded signal is pre-processed using a pre-emphasis filter. Pre-emphasis increases the magnitude of the higher frequencies as compared to the lower frequencies in order to equalize the speech frequency characteristics. For this purpose, a simple first order filter is used:

$$H(z) = 1 - 0.97z^{-1} \tag{1}$$

Block diagram representing general speaker recognition system is shown in figure 2.

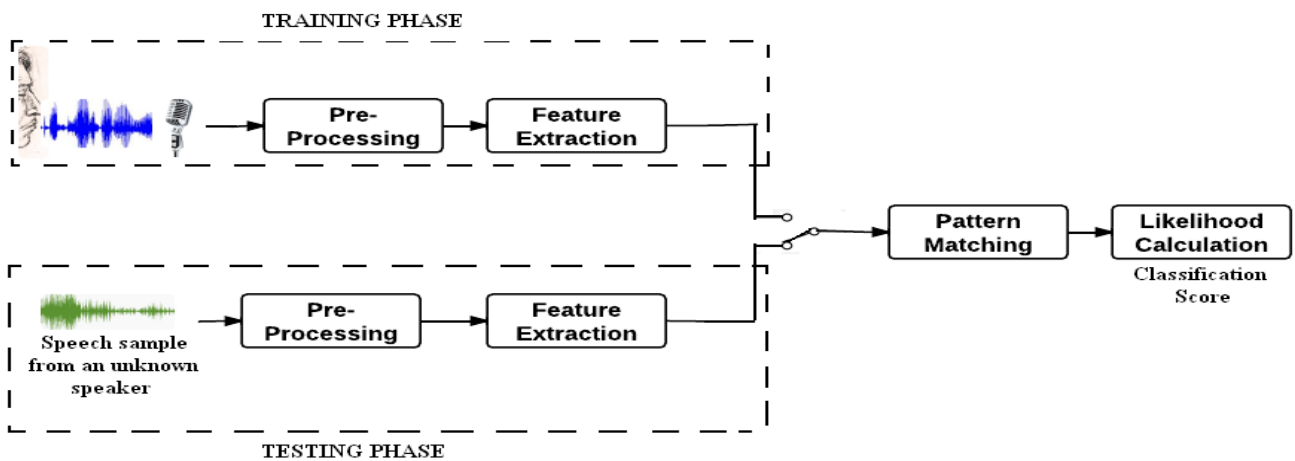

Fig.2 Block Diagram of a Speaker Recognition system

Both the training and the testing phase involve extraction of speech features also known as the enrollment phase followed by pattern matching process. During the enrollment process, speech samples from a number of speakers are recorded and speaker specific information is extracted using one out of the several methods available to generate feature vectors corresponding to an individual's voice model at a reduced data rate. Pattern matching phase, essentially involves two steps namely, pattern training and pattern comparison. For a speaker identification system, pattern comparison of the unknown speaker's voice model is performed with the voice model of the all the known reference speakers. The unknown speaker is identified as that reference speaker whose voice model best matches with the model of unknown utterance. For speaker verification applications, unknown speech template is only matched with the voice print of the claimed reference speaker. The unknown speaker is validated as the claimed speaker only if the classification score of the pattern matching is greater than some threshold value.
The performance of speaker identification system decreases with increasing population size while the performance of speaker verification system remains fairly constant with respect to the population size and it depends upon the extensiveness of the training phase. [2]

## 2.1 Feature Extraction

The mechanism of speech feature extraction reduces the dimensionality of the input signal by eliminating the redundant information while maintaining the discriminating capability of the signal [13]. Given the data of speech samples, a variety of auditory features are computed for each input set which constitute the feature vector. The present research proposes WP-MFC feature extraction approach.

## 2.1.2 Discrete Wavelet and Wavelet Packet Transform:

**Discrete Wavelet Transform**

To obtain an intuitive understanding of multiresolution we start with an example. Consider a landscape scene from an airplane high above the ground. Only major features like mountains, rivers and big geological structures can be seen. We would call it coarse or low level resolution of landscape. Moving in closer more details can be seen like houses roads etc. Moving even closer than humans, cars can be seen. Each stage of resolution is considered as space and each resolution space is a subset of a resolution scale of higher level. Thus, we can say that the 'mountain space' is a subset of 'house space' which is a subset of 'human space' and ultimately which is a subset of 'atomic space'. Hence multi-resolution signal decomposition of L2(R) is a nested series of closed subspaces Sj with the following properties [16].

$$\{0\} \subset S_\infty \subset ... \subset S_2 \subset S_1 \subset S_0 \subset S_{-1} \subset S_{-2} \subset .... \subset L^2(R) \tag{4}$$

For our finite approximations we always set the lowest detail to zero.

$$\lim_{j \to -\infty} S_j = U_{j \to -\infty}^\infty S_j = L^2(R) \tag{5}$$

$$\bigcap_{j \to -\infty}^\infty S_j = \{0\} \tag{6}$$

$$f(x) \in S_0 \leftrightarrow f(\frac{x}{2^j}) \in S_j \tag{7}$$

and there exists a function $\phi(x)$ belonging to L2(R) whose integer translates
$$\{\varphi(x-k) \mid k = ...,-2,-1,0,1,2,...\} \tag{8}$$

is an orthonormal basis3 in L2(R). We also say that $\phi(x)$ generates multiresolution signal decomposition {Sj}. The term fj(x) is the representation of f on the scale space Sj and contains all details of f(x) up to finer resolution level j. Equation (4) shows that functions of Sj could be represented as a linear combination of functions of S0. Equation (5) says that the signal approximation fj(x) from Sj converges to an original signal f (x) when j→−∞ (the precision becomes finer and finer) and when j→+∞ (the precision becomes coarser and coarser). Equation (6) states that we lose all the details when j→∞. Equation (7) states that Sj is the 2j scale version of S0 by changing scale and Sj is spanned by the scaled functions.

$$\{\varphi_{j,k}(x) = (\frac{1}{\sqrt{2^j}}).\varphi(\frac{x-2^j k}{2^j}) \mid k = ...,-2,-1,0,1,2,...\} \tag{9}$$

that is, each element f (x) in Sj( j fixed) can be written in the following form:

$$f(x) = \sum_{k=-\infty}^{\infty} C_{j,k} \varphi_{j,k}(x) \tag{10}$$

The function generating a Multi Resolution Signal Decomposition (MRSD) is called a scaling function. As any scaling function $\varphi(x) \in S_0 \cup S_{-1}$, $\varphi$ (x) can satisfy a scaling equation, that is, there exists a sequence H = {$h_k$| k = . . . ,−2,−1, 0, 1, 2, . . . } of real numbers with, just like the relation of a length on different scales,

$$\varphi(x) = \sqrt{2} \sum_{k=-\infty}^{\infty} h_k \varphi(2x - k) \tag{11}$$

Where the coefficient $h_k$ can be represented as,

$$h_k = <\varphi(x), \sqrt{2}\varphi(2x-k)> \tag{12}$$

For every MRSD there exists a wavelet function ψ(x) whose scaled and translated versions of

$$\{\psi_{j,k}(x) = (\frac{1}{\sqrt{2^j}})\psi(\frac{x-2^j k}{2^j}) \mid k = ...,-2.-1.0,1,2,....\} \tag{13}$$

The wavelet and scaling functions are related to each other, such that

$$S_{j-1} = S_j \cup W_j \tag{14}$$

which means that any function in $S_{j-1}$ can be split into two orthogonal parts, one in $S_j$ and the other in $W_j$ [17].

**The Fast Wavelet Transform**

For discrete wavelet transform we have,

$$f[n] = (\frac{1}{\sqrt{M}})\sum_k W_\Phi[j_0,k]\varphi_{j_0,k}[n] + (\frac{1}{\sqrt{M}})\sum_{j=j_0}^{\infty}\sum_k W_\psi[j,k]\psi_{j,k}[n] \tag{15}$$

Here f[n], φ$_{j0,k}$[n] and ψ$_{j,k}$[n] are discrete functions defined in [0,M-1], a total of M points. Now,

$$W_\varphi[j_0,k] = (\frac{1}{\sqrt{M}})\sum_n f[n]\varphi_{j_0,k}[n] \tag{16}$$

$$W_\Psi[j,k] = (\frac{1}{\sqrt{M}})\sum_n f[n]\psi_{j,k}[n] \text{ for } j \geq j_0 \tag{17}$$

(16) are called approximation coefficients while (17) are called detailed coefficients.
The coefficients which are defined for a particular wavelet by the equations (16) and (17), if could be found by another without knowing the scaling and dilation version of scaling and wavelet function. Then, the computation time can be reduced. Now,

$$\varphi_{j,k}[n] = 2^{j/2}\varphi[2^j n - k] = \sum_{n'} h_\varphi[n']\sqrt{2}\varphi[2(2^j n - k) - n'] \tag{18}$$

Let n'=m-2k, we have

$$\varphi_{j,k}[n] = \sum_m h_\varphi[m-2k]\sqrt{2}\varphi[2^{j+1}n - m] \tag{19}$$

Combining with equation (16) we get,

$$W_\varphi[j,k] = h_\varphi[-n]*W_\varphi[j+1,n]|_{n=2k,k\geq 0} \tag{20}$$

Similarly, combining with equation (17) we get,

$$W_\varphi[j,k] = h_\varphi[-n]*W_\varphi[j+1,n]|_{n=2k,k\geq 0} \tag{21}$$

This algorithm is "fast" because one can find the coefficients level by level rather than directly using (16) and (17) to find the coefficients. This algorithm was first proposed in [18]. This algorithm can be pictorially represented as

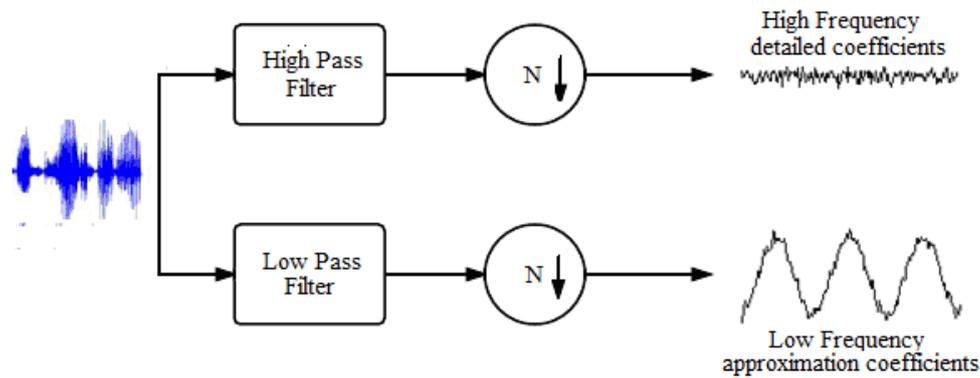

Fig.6 Fast Wavelet Transform Algorithm

**Wavelet Packet Transform**

In the DWT decomposition, to obtain the next level coefficients, approximation coefficients of the current level are split by filtering and downsampling. With the wavelet packet decomposition, the detail coefficients are also split by filtering and down sampling. The splitting of both the low and high frequency spectra results in a full binary tree shown in Figure.8 and a completely evenly spaced frequency resolution as illustrated in Figure 7.

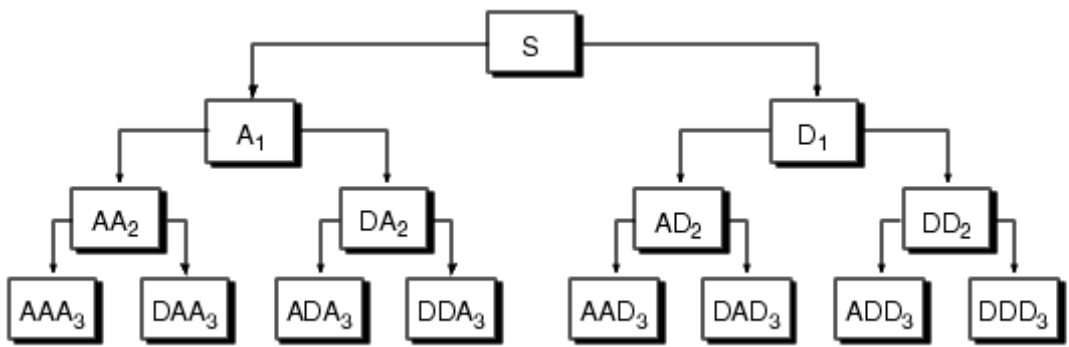

Fig.7 Wavelet packet decomposition tree

# 3 Classification Method

## 3.1 Hidden Markov Model

Hidden Markov Model (HMM) [19] [20] springs forth from Markov Processes or Markov Chains. It is a canonical probabilistic model for the sequential or temporal data it depends upon the fundamental fact of real world, "Future is independent of the past but driven by the present". The HMM is a doubly embedded stochastic process, where final output of the system at a particular instant of time depends upon the state of the system and the output generated by that state.

There are two types of HMMs: Discrete HMMs and Continuous Density HMMs. These are distinguished by the type of data that they operate upon. Discrete HMMs (DHMMs) operate on quantized data or symbols, on the other hand, the continuous density HMMs (HMMs) operate on continuous data and their emission matrices are the distribution functions. The basic notations of HMM are as shown in Table I.

TABLE I
BASIC NOTATIONS

| Variable | Notations |
|---|---|
| Number of States | $N$ |
| Number of observation symbols per state | $M$ |
| Observation symbols | $V\{v_1, v_2, \ldots, v_M\}$ |
| Observation Sequence | $O\{O_1, O_2 \ldots, O_T\} \in X$ {discrete value, real value} |
| State Sequence | $Q\{q_1, q_2, \ldots, q_T\}$ |
| Transition Matrix | $A = \{a_{i,j}\} = P(q_{t+1} = S_j | q_t = S_i) \; 1 \leq i, j \leq N$ |
| Emission Matrix | $b_i(k) = P(v_k \text{ at } t | q_t = S_i) \; 1 \leq i \leq N; \; 1 \leq k \leq M$ |
| Initialization Matrix | $\pi(i) = P(q_1 = S_i) \; 1 \leq i \leq N$ |
| Space of all state sequence of length $T$ | $\varrho$ |
| Mixture component for each state at each time | $m\{m_{q_1}, m_{q_2}, \ldots m_{q_T}\}$ |
| Mixture component (i state and $l$ component) | $c_{il}, \mu_{il}, \Sigma_{il}$ |
| Model of the System | $\lambda(A, B, \pi)$ |
| Parameter for Maximum Likelihood estimation | $\lambda_{ML}$ |
| EM Auxiliary Function | $Q(\lambda, \lambda^{(i-1)})$, where superscript for iteration (i-1) |

There are three major design problems associated with an HMM outlined here:
1) Given the Observation Sequence {O1, O2, O3,.., OT} and Model λ(A, B, π), the first problem is the computation of probability of the observation sequence P (O|λ).
2) The second problem finds the most probable state sequence Q {q1, q2,… ,qT}
3) The third problem is related to the choice of the model parameters λ (A, B, π), such that the Probability of the Observation sequence, P (O|λ) is the maximum.
The solution to the above problems emerges from three algorithms: Forward, Viterbi and Baum-Welch [19]

### 3.1.1 Continuous density HMM

Let O = {O1,O2,..,OT} be the observation sequence and Q {q1, q2,… ,qT} be the hidden state sequence. Now, we briefly define the Expectation Maximization (EM) algorithm for finding the maximum-likelihood estimate (MLE) of the parameters of a HMM given a set of observed feature vectors. The Q function is generally defined as

$$Q(\lambda, \lambda^{(i-1)}) = \sum_{q \in \varrho} \log P(O, Q | \lambda) P(O, Q | \lambda^{(i-1)}) \qquad (22)$$

The E – step and the M-step of EM algorithm are as follows:

*E Step:* $Q(\lambda, \lambda^{(i-1)}) =$

$$\sum_{Q \in \varrho} \sum_{m \in M} \log P(O, Q, m | \lambda) P(O, Q, m | \lambda^{(i-1)}) \qquad (23)$$

*M Step:*

$$\lambda^{(i)} = \arg\max_{\lambda}[Q(\lambda, \lambda^{(i-1)})] + constainment \qquad (24)$$

Final Result: when i→∞, λ$^{(i-1)}$→λ$_{ML}$

The optimized equations for the parameters of the mixture density are. [20]

$$\mu_{il} = \frac{\sum_{t=1}^{T} O_t P(q_t = i, m_{q_t t} = 1 | O, \lambda^{(i-1)})}{\sum_{t=1}^{T} P(q_t = i, m_{q_t t} = 1 | O, \lambda^{(i-1)})} \qquad (25)$$

$$\sum_{il} = \frac{\sum_{t=1}^{T}(O_t - \mu_{il})(O_t - \mu_{il})^T P(q_t = i, m_{q_t t} = 1 | O, \lambda^{(i-1)})}{\sum_{t=1}^{T} P(q_t = i, m_{q_t t} = 1 | O, \lambda^{(i-1)})} \qquad (26)$$

$$c_{il} = \frac{\sum_{t=1}^{T} P(q_t = i, m_{q_t t} = 1 | O, \lambda^{(i-1)})}{\sum_{t=1}^{T} \sum_{l=1}^{M} P(q_t = i, m_{q_t t} = 1 | O, \lambda^{(i-1)})} \qquad (27)$$

## 4 Motivation

Speech is a "Quasi-stationary" signal. MFCC utilizes short time Fourier Transform which provides information regarding the occurrence of a particular frequency at a particular time instant with a limited precision, with the resolution according to the Heisenberg Uncertainty principle dependent on the size of the analysis window.

$$Time * Frequency = \Delta t \Delta f \geq \frac{1}{4\pi}$$

Narrower windows provide better time resolution while wider ones provide better frequency resolution [21]. Even though STFT tries to strike a balance between the time and frequency resolution, it is admonished primarily as it keeps the length of the analysis window fixed for all frequencies resulting in uniform-partition of the time-frequency plane.

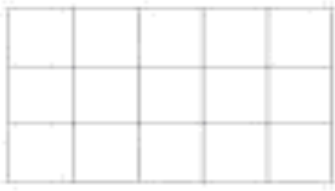
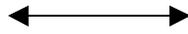
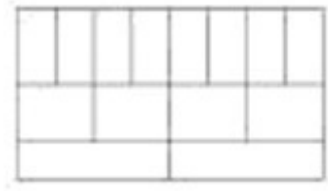

Fig.9 Time-frequency plane uniformly partitioned in STFT

Fig.10 Time-frequency plane non-uniformly spaced (constant area) in wavelet transform

Speech signals require a more flexible multi-resolution approach where window length can be varied according to the requirement to cater better time or frequency resolution. Wavelet Packet Transform (WPT) offers a remedy to this difficulty by providing well localized time and frequency resolutions as shown in figure 10 [22]. Further, multi-resolution property of WPT makes it more robust in noisy environment as compared to single-resolution techniques. Also, in this work WPT has been utilized instead of wavelet transform because with the increase of spatial resolution, frequency resolution decreases, which is a disadvantage of wavelet transform. By further dividing frequency domain, wavelet packet transform can overcome the deficiency of wavelet transform and has better time-frequency characteristic. But, WPT increases the computational burden and is time consuming and conventional wavelet packet transform mechanisms do not warp the frequencies according to the human auditory perception system. Since wavelet packet coefficients does not exist completely in either time or the frequency domain but, in a mixed domain of time and frequency, therefore the coefficients obtained by wavelet packet transform can't be warped to Mel Frequency Cepstrum. Hence, in the present work we suggest an approach utilizing the advantages of the Mel Scale and multi-resolution WPT to generate WP-MFC feature vectors and perform its comparative study with the conventional MFCC for the task of speaker recognition.

# 5 Proposed Method

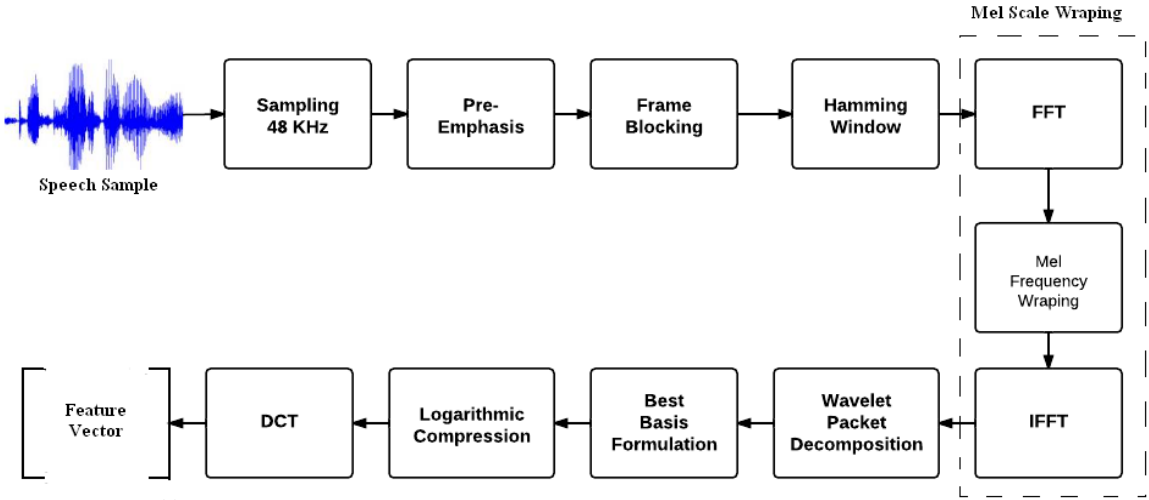

Fig.11 Block diagram representation of proposed method

The analytical steps followed for feature extraction as depicted in Fig. 11 are as stated:

i. The raw input speech signal was primarily sampled at 48 kHz in order to further process it.
ii. The sampled speech signal was next passed through a hamming window. A frame size of 25 milliseconds and a skip rate of 10 milliseconds were chosen to accommodate for the best continuity.
iii. A pre-emphasis filter as described by equation (12) was next exercised in order to improve the overall signal-to-noise ratio. A rectangular Hamming window was deployed for framing as depicted in equation (1).
iv. The conversion of the signal from time to frequency domain was obtained by applying Fast Fourier Transform (FFT). The resultant signal in frequency domain was Mel-Wrapped using 40 Triangular Mel Filter Banks. Afterwards, signal was converted back to time domain by applying Inverse Fast Fourier Transform (IFFT) for further processing of signal.
v. Next, a daubechies4 (D4) wavelet was employed to carry out the Wavelet Packet Decomposition. For a full j=7 level decomposition, the wavelet packet transform corresponds to a maximum frequency of 31.25 Hz, thus giving us 128 sub-bands.
vi. Out of 128 frequency sub-bands, 35 frequency sub-bands were used for further processing since higher frequency coefficient contained insignificant amount of energy and first 35 coefficients represented 99.99% of the total energy output. The energy in each band was then evaluated, and was divided by the total number of coefficients present in that particular band. In particular, the sub band signal energies were computed for each frame as,

$$E_j = \frac{\sum_{j=1}^{N_j}[W_j^p f(i)]^2}{N_j}, j = 1,...,35 \tag{28}$$

vii. Lastly, a logarithmic compression was performed and a Discrete Cosine Transform (DCT) was applied on the logarithmic sub-band energies to reduce dimensionality:

$$F(i) = \sum_{n=1}^{B} \log_{10} E_n Cos(\frac{i(n-1/2)}{B}), i = 1,...,r. \qquad (29)$$

## 5.1 Speaker recognition

### 5.1.1 Speaker Identification:

After extracting the features we have used HMM or single state HMM called GMM (Gaussian Mixture Model) for the identification. The whole procedure is as explained in Fig.12. Having the WP-MFC Feature from the speech signals, CDHMMs are trained for each speaker using Baum Welch (BM) algorithm which gives the parameters of the corresponding CDHMMs. Now the identification process can be described as follows: Given a test vector 'X' the log-likelihood of the trained batches with respect to their HMM models '$\lambda$' is computed as logP(X|$\lambda$).

From 'N' HMMs $\{\lambda_1, \lambda_2, ........\lambda_N\}$ corresponding to 'N' speakers, the speaker can be identified with a test sequence using:

$$P(X | \lambda_{required}) = F[P(X | \lambda_1),..........P(X | \lambda_N)] \qquad (30)$$

Where $F()$ is the maximum of the likelihood values of the model $(\lambda_1, \lambda_2,........\lambda_N)$. The model corresponding to the highest Log-Likelihood value is selected as the identified speaker.

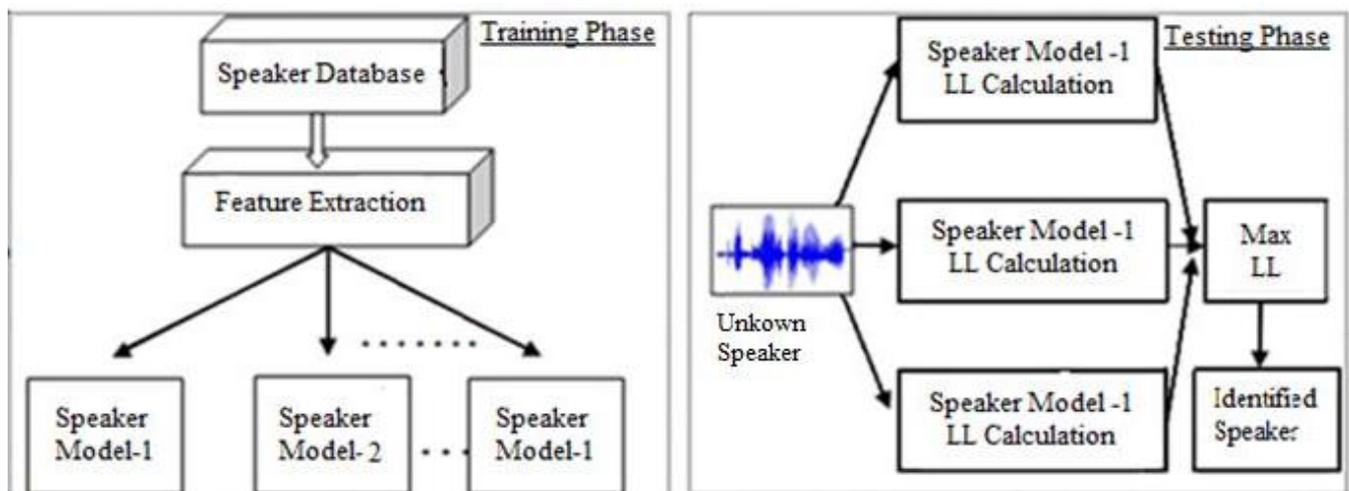

Fig.12 Procedure for the Speaker Identification

### 5.1.2 Speaker Verification:

After extracting the features we have used HMM or single state HMM called GMM (Gaussian Mixture Model) for the task of verification. The procedure is as explained in Fig.13. Having the WP-MFC Feature from the speech signals, CDHMM are trained for a particular speaker using Baum Welch (BM) algorithm using speech samples of the speaker which gives the parameters of the corresponding CDHMM and threshold value was obtained by using different speech samples of the same speaker and calculating the likelihood from the CDHMM. Now the verification process can be described as follows: Given a test vector 'X' the log-likelihood of the trained batch with respect to the HMM model '$\lambda$' is computed as logP(X|$\lambda$).

From the trained HMM corresponding to the true speaker, the speaker's identity can we validated by comparing the obtained likelihood with the threshold obtained during the testing phase.

$$if\,(\log P(X\mid\lambda)\geq threshold\,) \tag{31}$$

A binary decision true or false is obtained using equation (34) where the speaker is validated as true speaker if the value of likelihood obtained for the test vector is greater than or equal to the threshold value and the speaker is rejected as an imposter if likelihood value is less than threshold.

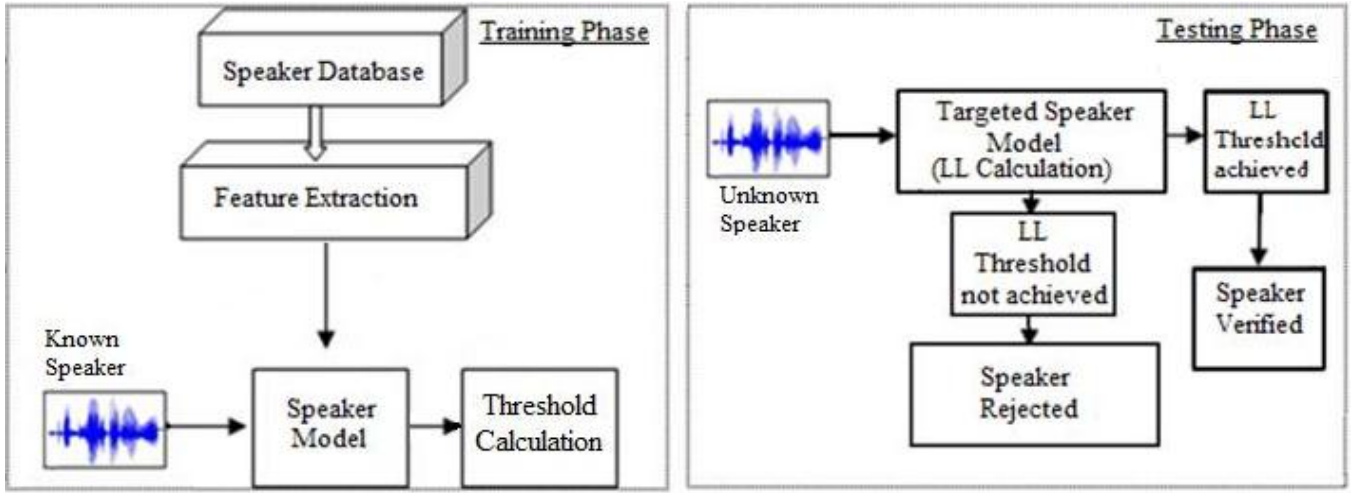

Fig.13 Procedure for the Speaker Verification

# 6 Results

For the task of speaker identification we acquired the database from [23]. The speakers were randomly chosen to validate the models on the real-time test speech signals. This database consists of 40 speakers with 10 voice samples each with sampling rate of 48000, 16 bit PCM and of mono-channel recording using different means of recording transducer from different regions of the world. We have used eight samples for the training (training dataset) and two for the testing (testing dataset).The dataset used for testing is mixed with noise signals at different SNRs (-5db, 5db, 10db, 20db). The noise signals are taken from NOISEX database [24]. For evaluating the performance of the proposed models on a set of 40 speakers, two type of noise: car noise and factory noise are used.

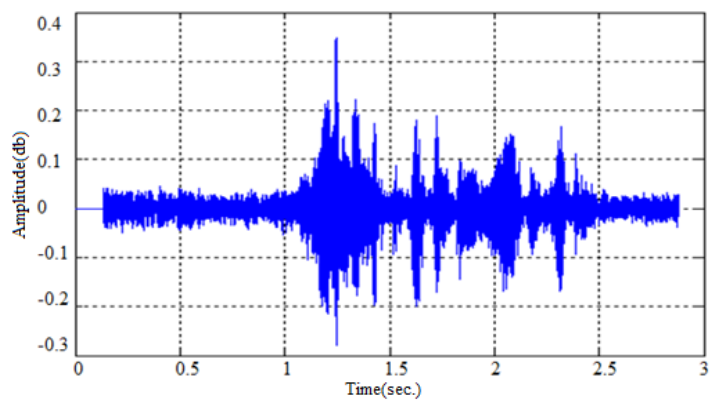

Fig. 14 A Sample Speech Signal

For the task of speaker verification we acquired a standard free database CSTR US KED TIMIT from the site [25].The database contains 453 utterances of a US male speaker. This database was collected at University of Edinburgh's Centre for Speech Technology Research [26]. For testing of the imposter speakers test samples are obtained from [23]. All test samples are different and distinct in content.

## 6.1 MFCC Data Description

The raw speech signal was sampled at 48Khz. Pre-emphasis filtering was performed in order to improve the overall signal-to-noise ratio. Next, a hamming window of 25msec duration was used for framing. Overlap of 15 milliseconds was chosen for eliminating the possibility of loss of signal. The speech samples were transformed to the frequency domain by performing FFT on them. The output spectrum was then Mel-Wrapped using 40 triangular filters yielding the Mel Frequency Wrapped Spectrum using 13 filter bank coefficients out of these 40. Next, we logarithmically compressed the signal followed by the Discrete Cosine Transform which provided us with the Mel Frequency Cepstral Coefficients.

TABLE III
MFCC FEATURE SAMPLES

|   | 1 | 2 | 3 | 4 | ... | ... | ... | 12 | 13 |
|---|---|---|---|---|-----|-----|-----|----|----|
| 1 | -15.8742 | -9.67495 | -1.68234 | -8.0967 | ... | ... | ... | 4.8634871 | -2.33387 |
| 2 | -16.3469 | -10.8389 | -3.59678 | -3.58626 | ... | ... | ... | 3.377781 | 3.000956 |
| 3 | -15.7171 | -11.0581 | -3.49327 | -1.16696 | ... | ... | ... | 5.189478 | -2.88547 |
| 4 | -15.2708 | -11.0081 | -1.03966 | -0.51895 | ... | ... | ... | 5.189478 | -2.91193 |
| ... | ... | ... | ... | ... | ... | ... | ... | ... | ... |
| ... | ... | ... | ... | ... | ... | ... | ... | ... | ... |
| ... | ... | ... | ... | ... | ... | ... | ... | ... | ... |
| ... | ... | ... | ... | ... | ... | ... | ... | ... | ... |
| 747 | 63.45957 | 3.109013 | 2.776829 | 4.691779 | ... | ... | ... | -1.47351 | 3.200046 |
| 748 | 62.79702 | 1.526184 | 1.602693 | 4.950456 | ... | ... | ... | -4.06394 | 1.990352 |

Each window has 13 coefficients (columns) which represent the content in the corresponding window of the sample. Numbers of windows (rows) vary with the length of the speech sample.

## 6.2 Wavelet Packet Based MFCC Data Description

The speech signal was sampled at 48 kHz. The signal was then passed through pre-emphasis filter in order to increase the overall signal to noise ratio. Sample was then framed using a hamming window of size 25 msec. so that every segment could harbor a number of samples divisible by $128(2^7)$ and with an overlap of 15 msec. in order to minimize signal loss. The resultant signal was then Mel-Wrapped by the already mentioned method. Next, 7 level wavelet packet decomposition was applied using 'Daubechie4' wavelet. For a full j=7 level decomposition, the wavelet packet transform dividing the frequency axis into 128 sub-bands of equal bandwidth, corresponding to a maximum frequency of 31.25 Hz. Out of these 128 sub-bands, 66 sub-bands were chosen after best basis formulation. After this, sub band signal energies were computed for each frame. Finally, a Discrete Cosine Transform was applied on the logarithmically compressed sub-band energies to reduce dimensionality.

TABLE IV

Table IV
WAVELET PACKET BASED MFCC FEATURE SAMPLES

|     | 1        | 2        | 3         | 4        | ... | ... | ... | 34       | 35       |
|-----|----------|----------|-----------|----------|-----|-----|-----|----------|----------|
| 1   | -50.8012 | 9.48821  | -9.34097  | -6.31279 | ... | ... | ... | -7.75253 | 27.67921 |
| 2   | -52.5288 | 8.006746 | -12.9264  | -2.59287 | ... | ... | ... | 0.456589 | 32.94345 |
| 3   | -51.1801 | 7.343666 | -10.6692  | -4.17875 | ... | ... | ... | -5.05226 | 29.86128 |
| 4   | -51.9652 | 11.0148  | -14.0033  | 2.05767  | ... | ... | ... | -18.1968 | 39.32132 |
| ... | ...      | ...      | ...       | ...      | ... | ... | ... | ...      | ...      |
| ... | ...      | ...      | ...       | ...      | ... | ... | ... | ...      | ...      |
| ... | ...      | ...      | ...       | ...      | ... | ... | ... | ...      | ...      |
| ... | ...      | ...      | ...       | ...      | ... | ... | ... | ...      | ...      |
| 760 | 43.38049 | 6.924977 | -8.86377  | -5.70881 | ... | ... | ... | -6.64844 | 24.12309 |
| 761 | 43.30697 | 7.602423 | -9.79684  | -4.50895 | ... | ... | ... | -9.45975 | 29.68451 |

Only the first 35 coefficients were evaluated since lower frequency components are highly dominant in a human speech signal. The length of the feature vector (i.e. the number of rows) depends on the length of the speech signal.

## 6.3 Speaker Identification Results

The following section gives a tabular as well as pictorial representation of the results we obtained for the task of speaker recognition. In order to find the best parameters for the CDHMMs i.e. the number of mixtures and (M) and number of states (Q), number of states (Q) were kept constant first and the number of mixtures (M) were varied in each case. The result obtained shows that the optimized number of mixtures is 13 and number of states is 1 (as shown in table V, VI and figure 15).

TABLE V
SPEAKER IDENTIFICATION RESULTS FOR NO. OF STATES (Q) = 1

| S.No. | GMM Parameters      |                   | Speakers Identified ( in % ) |              |
|-------|---------------------|-------------------|------------------------------|--------------|
|       | No. of mixtures (M) | No. of states (Q) | WP-MFC features              | MFCC features |
| 1.    | 11                  | 1                 | 97.5                         | 95.0         |
| 2.    | 12                  | 1                 | 97.5                         | 95.0         |
| 3.    | 13                  | 1                 | 97.5                         | 95.0         |
| 4.    | 14                  | 1                 | 97.5                         | 95.0         |
| 5.    | 15                  | 1                 | 97.5                         | 95.0         |

TABLE VI
SPEAKER IDENTIFICATION RESULTS FOR NO. OF STATES (Q) = 2

| S.No. | HMM Parameters | | Speakers Identified ( in % ) | |
|---|---|---|---|---|
| | No. of mixtures (M) | No. of states (Q) | WP-MFC features | MFCC features |
| 1. | 11 | 2 | 90.0 | 85.0 |
| 2. | 12 | 2 | 97.5 | 95.0 |
| 3. | 13 | 2 | 97.5 | 95.0 |
| 4. | 14 | 2 | 90.0 | 95.0 |
| 5. | 15 | 2 | 90.0 | 97.5 |

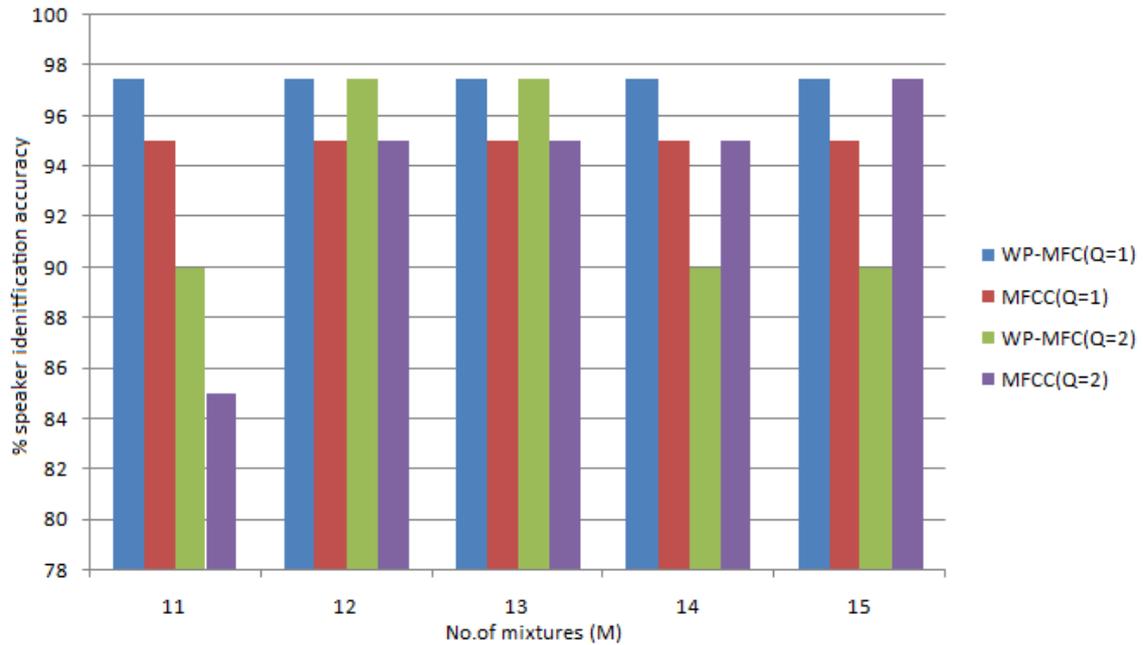

Fig. 15 Pictorial representation of speaker identification

The results obtained thus shows that GMM is a better classifier for the task of speaker identification with optimized no. of mixtures (M) are 13 and no. of states (Q) is 1. To check the noise robustness of WP-MFC features standard testing noises (Car and Factory) are added to test samples and then the noise corrupted test samples are evaluated for both MFCC and WP-MFC features. The result obtained demonstrates the noise robustness of the WP-MFC features (as shown in table VII, VIII and figure 16).

TABLE VII
SPEAKER IDENTIFICATION RESULTS FOR CAR NOISE OF DIFFERENT SNRS

| S.No. | Noise Level | GMM Parameters | | Speakers Recognized (in %) | |
|---|---|---|---|---|---|
| | | No. of mixtures(M) | No. of states (Q) | WP-MFC features | MFCC features |
| 1. | Clean Speech | 13 | 1 | 97.5 | 95.0 |
| 2. | 20dB | 13 | 1 | 97.5 | 92.5 |
| 3. | 10dB | 13 | 1 | 77.5 | 52.5 |
| 4. | 5dB | 13 | 1 | 60 | 37.5 |
| 5. | -5dB | 13 | 1 | 20 | 7.5 |

TABLE VIII
SPEAKER IDENTIFICATION RESULTS FOR FACTORY NOISE OF DIFFERENT SNRS

| S.No. | Noise Level | GMM Parameters | | Speakers Recognized (in %) | |
|---|---|---|---|---|---|
| | | No. of mixtures(M) | No. of states (Q) | WP-MFC features | MFCC features |
| 1. | Clean Speech | 13 | 1 | 97.5 | 95.0 |
| 2. | 20dB | 13 | 1 | 35.0 | 32.5 |
| 3. | 10dB | 13 | 1 | 15.0 | 15.0 |
| 4. | 5dB | 13 | 1 | 7.5 | 5.0 |
| 5. | -5dB | 13 | 1 | 5.0 | 5.0 |

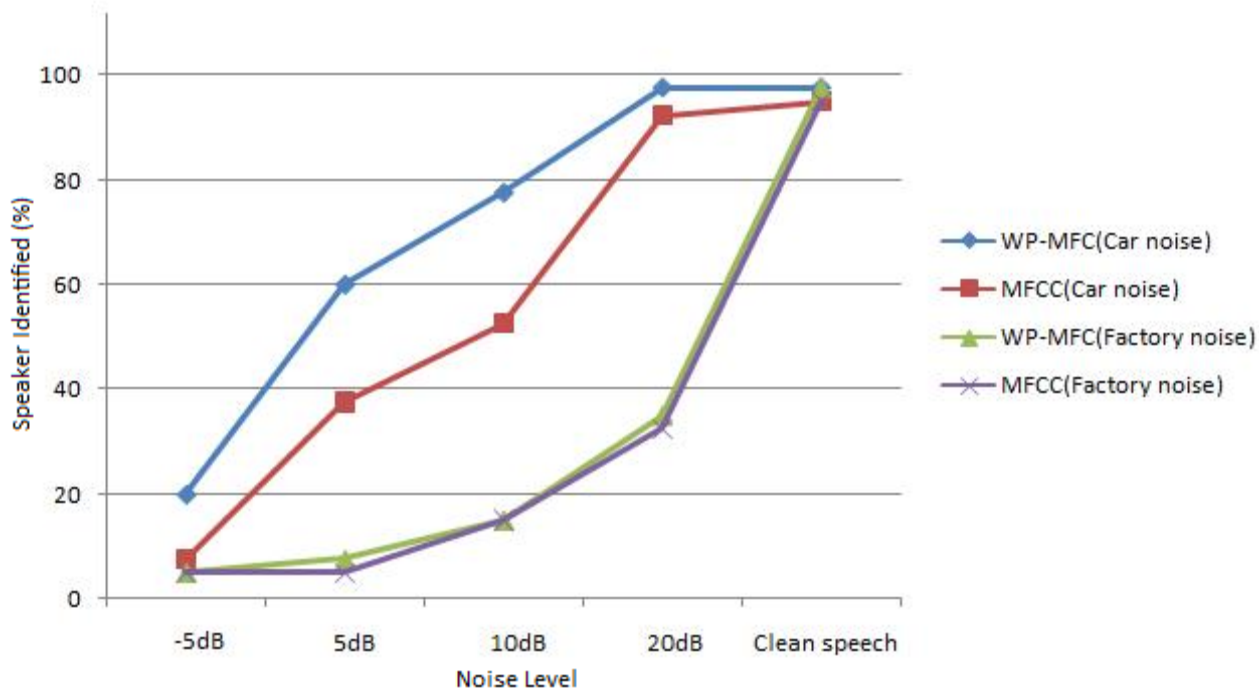

Fig.16 Pictorial representation of identification results for noisy test samples

## 6.3 Speaker Verification Results

For the task of speaker verification GMM was used as classifier with 13 mixtures (M) and 1 state (Q). The result obtained is show in table IX.

TABLE IX
SPEAKER VERIFICATION RESULTS

| S.No. | GMM Parameters | | Percentage Accuracy | | False Acceptance Percentage | |
|---|---|---|---|---|---|---|
| | No. of Mixtures (M) | No. of States (Q) | WP-MFC features | MFCC features | WP-MFC features | MFCC features |
| 1 | 13 | 1 | 96.5 | 93.5 | 3 | 4.5 |

# 7 Conclusions

Speaker Recognition is the use of machine to recognize a speaker from the spoken words. In this paper,

we introduced a robust feature extraction technique for deployment with speaker identification system. These new feature vectors termed as Wavelet Packet based Mel frequency Cepstral (WP-MFC) Coefficients offer better time and frequency resolution. HMM and GMM were used to classify the acoustic data. Experimental results of the comparison between the performance of the proposed feature vectors and MFCC reveal the real life effectiveness of the proposed method. For the task of speaker identification both in uncorrupted and noise corrupted testing signals proposed WP-MFC Features demonstrated better results than MFCC Features and better results were obtained with GMM as classifier than HMM, giving us the optimum parameters of the GMM i.e. no. of mixtures (M) and no. of states (Q). For the task of speaker verification better results were obtained with WP-MFC Features and with GMM as classifier using the optimum parameters obtained. No significant improvement was seen with increasing no. of states (Q) and no. of mixtures (M) in HMM.